\documentclass[reprint,english,aip,showpacs]{revtex4-1}
\pdfoutput=1
\usepackage{amsmath}
\usepackage{color}
\usepackage{graphicx}
\usepackage{epstopdf}
\usepackage{babel}

\begin{document}

\title{Diffraction at GaAs/Fe$_3$Si core/shell nanowires: the formation of nanofacets}
\author{B.~Jenichen}
\email{bernd.jenichen@pdi-berlin.de}
\author{M.~Hanke}
\author{M.~Hilse}
\author{J.~Herfort}
\author{A.~Trampert}
\address{Paul-Drude-Institut f\"ur Festk\"orperelektronik,
Hausvogteiplatz 5--7,D-10117 Berlin, Germany}
\author{S.~C.~Erwin}
\address{Center for Computational Materials Science,
Naval Research Laboratory,
Washington DC, 20375, USA}

\date{\today}

\begin{abstract}
GaAs/Fe$_{3}$Si core/shell nanowire structures were fabricated by molecular-beam epitaxy on oxidized Si(111) substrates and investigated by synchrotron x-ray diffraction. The surfaces of the Fe$_3$Si shells exhibit nanofacets. These facets consist of well pronounced Fe$_3$Si\{111\} planes. Density functional theory reveals that the Si--terminated Fe$_3$Si\{111\} surface has the lowest energy in agreement with the experimental findings. We can analyze the x-ray diffuse scattering and diffraction of the ensemble of nanowires avoiding the signal of the substrate and poly-crystalline films located between the wires. Fe$_3$Si nanofacets cause streaks in the x-ray reciprocal space map rotated by an azimuthal angle of 30~$^\circ$ compared with those of bare GaAs nanowires.
In the corresponding TEM micrograph the facets are revealed only if the incident electron beam is oriented along [1$\overline{1}$0] in accordance with the x-ray results.
Additional maxima in the x-ray scans indicate the onset of chemical reactions between Fe$_{3}$Si shells and GaAs cores occurring at increased growth temperatures.
\end{abstract}

%%Shell growth  at a substrate temperature of $T_{S}$~=~200~$^\circ$C leads to nanofacetted  Fe$_{3}$Si shells.

%%\pacs{61.05.cp,68.37.Lp,68.70.+w,68.55.ag,81.15.Hi}
%%61.05.cp	X-ray diffraction, 68.37.Lp TEM, 68.35.bd	Structure of surfaces and interfaces, Metals and alloys, 81.15.Hi	Molecular, atomic, ion, and chemical beam epitaxy

%%61.05.cp xrd, 68.37.Lp TEM, 68.55.ag semiconductors,68.70.+w, whiskers


\maketitle

\section{Introduction}
 Semiconductor/ferromagnet core/shell nanowires (NWs) have gained a lot of interest recently.\cite{Hilse2009,Rudolph2009,Rueffer2012,Dellas2010,Tivakorn2012,Yu2013,hilse2013,jenichen2015} The cylindrical shape of the ferromagnet in such core/shell NWs causes a magnetization along the wire axis, i.e. perpendicular to the substrate surface.\cite{hilse2013} The lattice matching of the binary Heusler alloy Fe$_{3}$Si and GaAs is a prerequisite for defect--free molecular beam epitaxy (MBE) of high quality hybrid structures.\cite{herfort03,Jenichen05,Herfort2005,Herfort2006,Herfort2006b}

% fig.1
\begin{figure*}
\centering

\includegraphics[width=6.0cm]{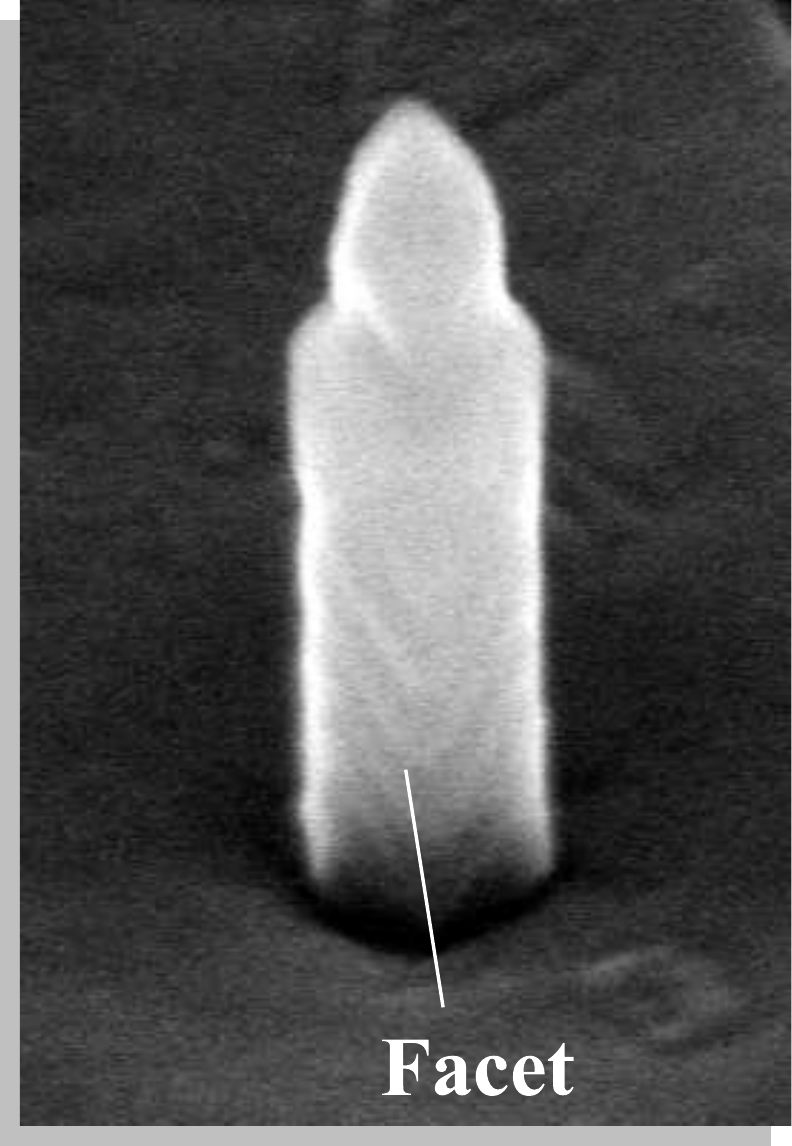}
\includegraphics[width=4.0cm]{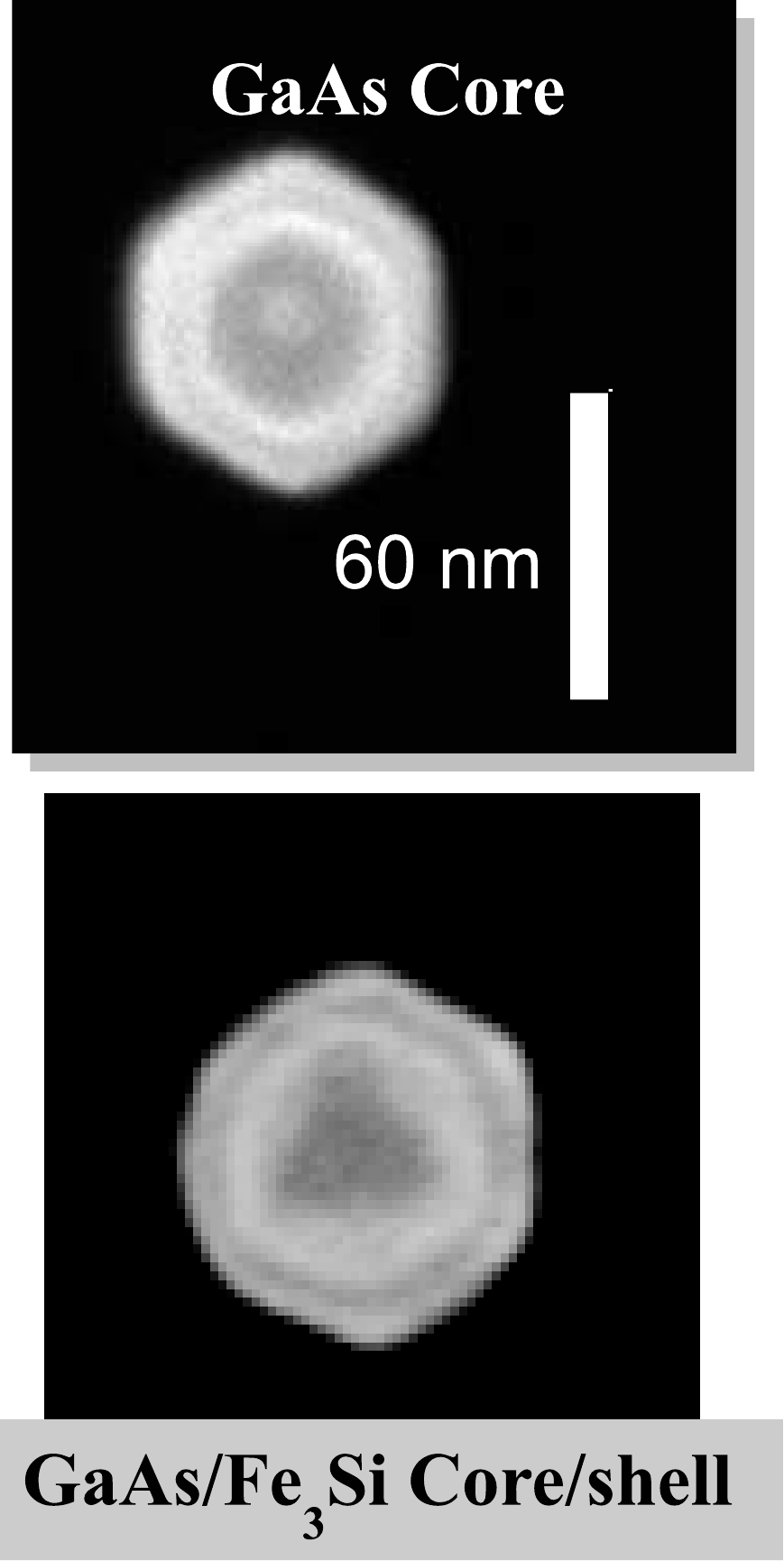}
\includegraphics[width=7.0cm]{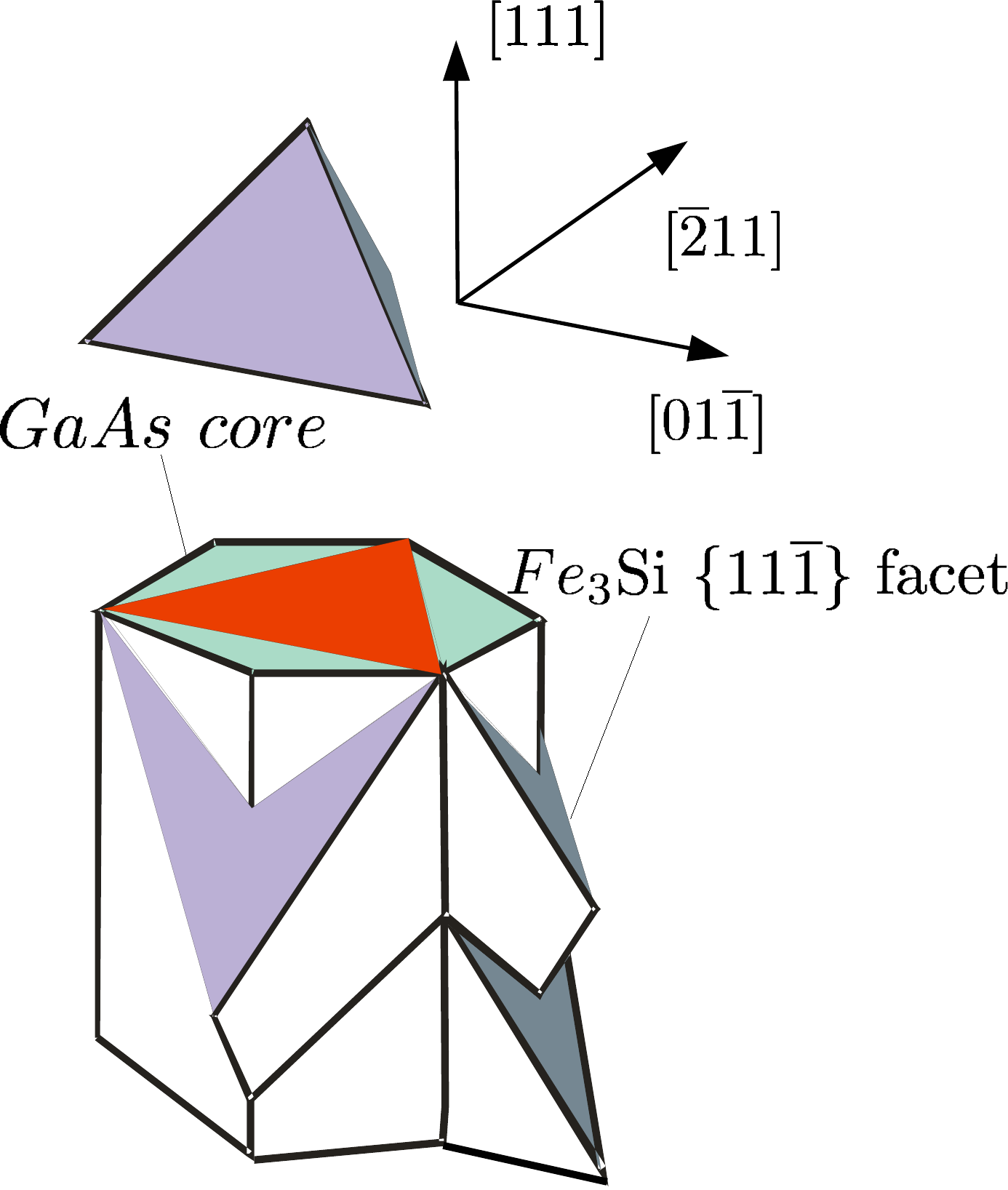}
\caption{(color online) SEM side view of  a core/shell NW  (left side) and  SEM top views of a core and core/shell NWs (center) grown at T$_S$~=~200~$^\circ$C. A sketch of the GaAs core and the Fe$_{3}$Si shells (right side) with tilted \{11$\overline{1}$\} facets is illustrating schematically the orientation relationship between the cores and the facets. In addition a tetrahedron consisting of \{111\} planes is given.
}
\label{fig:sketch_facets}
\end{figure*}

The nucleation of a film on a substrate can be considered in a similar manner as wetting or nonwetting of liquids.\cite{Bauer1958}
If the film wets the substrate it will grow layer-by-layer in  Frank van der Merwe growth mode. In this case the surface of the film is simply repeating the surface of the substrate. In  the non--wetting case
 three-dimensional (3D) islands will form in the Volmer-Weber (VW) growth mode, similar to liquid droplets.  Fe$_{3}$Si is lattice matched to GaAs. Nevertheless we observed the strain--free VW island growth mode of Fe$_{3}$Si on GaAs.\cite{kag09} Thus initially isolated islands can adopt their equilibrium shape before coalescence. New facets may arise in such a way.
Similarly small particles of Au on a MgO substrate e.g. adopt the equilibrium shape with $\{111\}$ and $\{100\}$ facets.\cite{marks1994} The equilibrium crystal shape of GaAs in an As-rich environment consists of $\{111\}$, $\{110\}$ and $\{100\}$ facets.\cite{moll1996} For the Ga-assisted catalyst-free growth of GaAs NWs by MBE in the [111] direction  only the $\{110\}$ sidewalls are present for geometrical reasons, as the growth rate of the NWs along [111] is dominating.\cite{Colombo2008} However for Au-induced NW growth $\{111\}$ nanofacets were found in the zincblende sections of the NWs.\cite{mariager2007}
MBE grown GaAs/AlAs core/shell NWs were decorated by Stranski Krastanov (SK) growth of InAs islands.\cite{Uccelli2010} These islands exhibit $\{111\}$ and $\{115\}$ facets and develop preferentially at the $\langle112\rangle$-oriented corners of the NWs. Ge islands on Si NWs exhibited  $\{111\}$, $\{110\}$ and $\{113\}$ facets as a result of a SK growth process.\cite{Pan2005} Ferromagnetic MnAs islands grow in a strain driven VW growth mode on semiconducting InAs NWs, where the lattice mismatch amounts up to 18 percent.\cite{Ramlan2006} In our system Fe$_{3}$Si/GaAs we can expect a strain--free VW island growth of Fe$_{3}$Si on GaAs\{110\}.\cite{kag09}

We recently demonstrated that GaAs/Fe$_{3}$Si core/shell NWs prepared by MBE show ferromagnetic properties with a magnetization oriented along the NW axis (i.e. perpendicular to the substrate surface). The properties of the NWs are determined by the growth temperature T$_S$ of the Fe$_{3}$Si.\cite{jenichen2014} For certain growth temperatures a coincidence of the core- and shell-orientations was observed by high-resolution transmission electron microscopy (HRTEM) and selected area diffraction (SAD) in the TEM.\cite{jenichen2015} Here the Fe$_{3}$Si growth is pseudomorphic on GaAs. The formation of facets at the Fe$_{3}$Si shell surface was detected as well. During TEM observations only a relatively small sample volume is probed. So it is reasonable to complement TEM measurements by an experiment that averages over many NWs. In this work, we investigate the Fe$_{3}$Si shells grown at different substrate temperatures T$_S$ and characterize the facets induced by strain--free VW growth of Fe$_{3}$Si shells using X-ray diffraction (XRD) in grazing incidence geometry, and scanning electron microscopy (SEM), together with HRTEM. With XRD we are facing the challenge to distinguish the signals of the core/shell NWs and the so-called parasitic metallic film which is unintentionally deposited between the NWs. The solution is the application of grazing incidence diffraction (GID) at zero incidence angle. In that way we can avoid the signals of the substrate and the parasitic layer.

On the other hand in the  present work we give a theoretical foundation to explain the experimental observations using density functional theory (DFT). We compare the energies of the different surfaces of the Fe$_{3}$Si shells.

\section{Theoretical Section}
The equilibrium state of the Fe$_{3}$Si surface as a function of composition is determined by the minimization of the grand potential $\Omega$,\cite{Landau1980,Vitos1998} where E is the energy of the crystal, S and T are the entropy and the temperature of the system, n$_{i}$ and $\mu_{i}$ are the numbers of atoms of sort i and the corresponding chemical potentials.
\begin{equation}
\Omega = E - TS - \sum n_{i}\mu_{i},
\end{equation}
This minimization is usually performed for the temperature T = 0.
In other words the stability of the surface (i.e. its crystallographic orientation and its surface termination or chemical configuration) is given by its surface energy per unit area $\gamma$, which is a free energy expressed with respect to the chemical potentials $\mu_{i}$, which represent reservoirs of the chemical species involved. This means:
\begin{equation}
\gamma\cdot A = E_{t} - n_{Fe}\mu_{Fe} - n_{Si}\mu_{Si},
\end{equation}
where A is the area of the surface unit cell, E$_{t}$ is its total energy, and n$_{Fe}$ and n$_{Si}$ are the numbers of Fe and Si atoms in the cell. We assume the surface to be in thermodynamic equilibrium with the bulk material. This implies that 3$\mu_{Fe}$  + $\mu_{Si}$ = g$_{Fe3Si}$, the energy per formula unit of bulk Fe$_{3}$Si.\cite{Martin2004,Qian1988} This constraint implies that g can be expressed as a function of just one independent variable e.g. $\mu_{Si}$. On the other hand the individual atomic chemical potentials can never exceed the energy of the condensed pure element, i.e. the energy per atom in bulk Fe and bulk Si. The three constraints jointly place an upper and lower limit on $\mu_{Si}$.
DFT\cite{Hohenberg1964,Kohn1965,Kohn1999,Hasnip2014,Jones2015} in the generalized gradient approximation\cite{Perdew1996} was applied in order to determine the surface energy of low index surfaces using the Vienna Ab Initio Simulation Package (VASP).\cite{Kresse1996A,Kresse1996B} The calculations were performed in a slab geometry using slabs from 12 to 16~${\AA}$ thick and a vacuum region of 10~${\AA}$. All atomic positions were relaxed except the innermost 3-4 layers, until the largest force component on every atom was below 0.05 eV~${\AA}^{-1}$. All of the calculations assumed a fully ordered ferromagnetic Fe$_{3}$Si lattice.

\section{Experimental Section}

GaAs/Fe$_{3}$Si core/shell NW structures were grown by MBE on Si(111) substrates.\cite{hilse2013,jenichen2015} 
The resulting NW structures were investigated by X-ray diffraction, SEM, and TEM.
The X-ray experiments were performed in grazing incidence geometry at the beamline BM~02 of the European Synchrotron Radiation Facility (ESRF) in Grenoble. The energy of the beam was 10~keV and the detector was an area detector S70 from Imxpad. The detector size is 560$\times$120 pixels of size 130$\times$130 $\mu$m$^{2}$ each. In order to minimize the penetration depth into substrate and parasitic film between the NWs a very small angle of incidence of 0.04~$^\circ$ was choosen.\cite{dosch86} In this way the signal of the ensemble of NWs was always dominating the diffraction pattern. This was checked by observation of the vanishing signal of the single-crystal Si substrate, which would dominate the diffraction curve at larger angles of incidence.
The TEM specimens are prepared as described in Ref.~\onlinecite{jenichen2015}.

\section{Results and discussion}

% fig.2
\begin{figure}[!t]
\includegraphics[width=9.0cm]{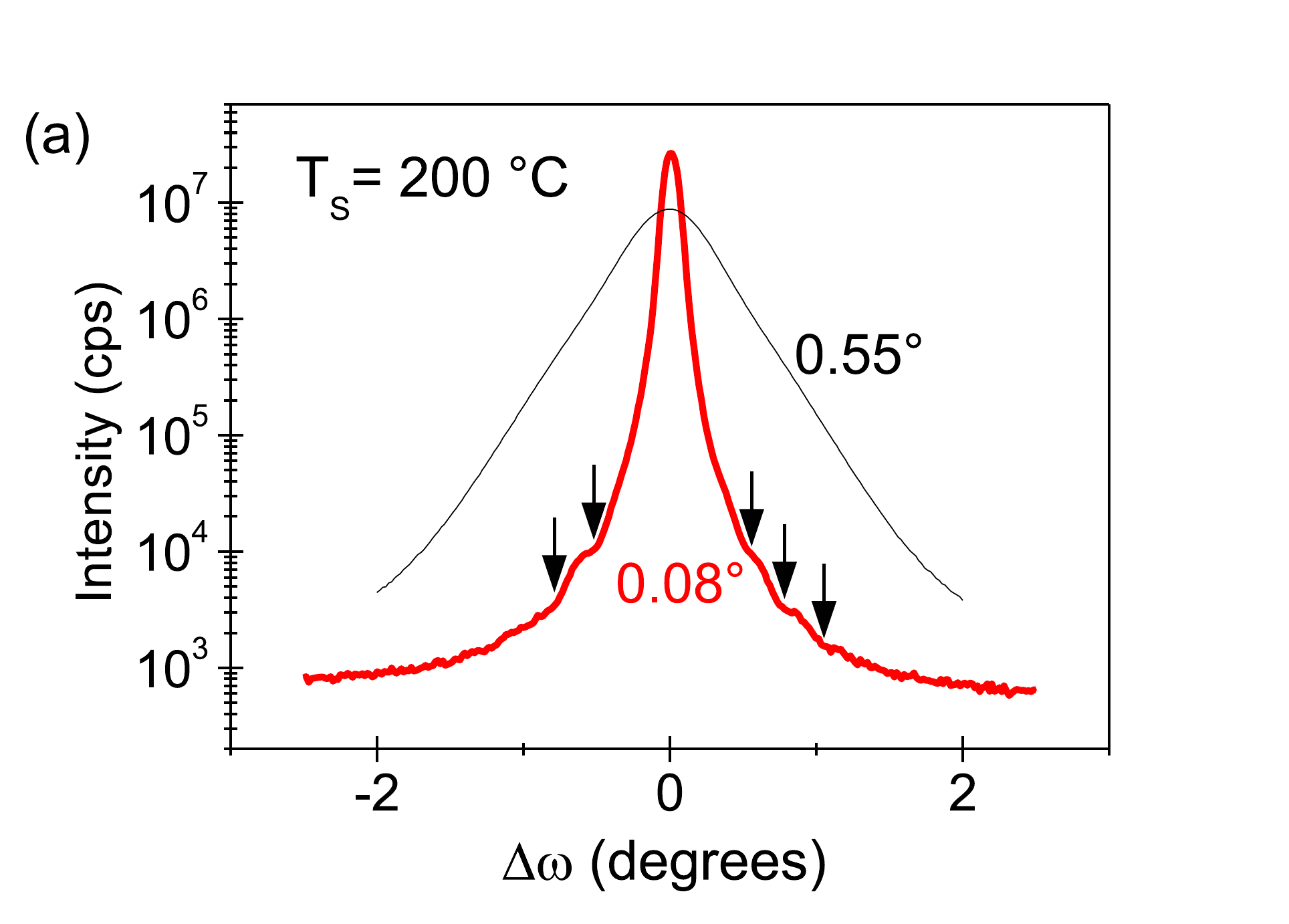}
\includegraphics[width=9.0cm]{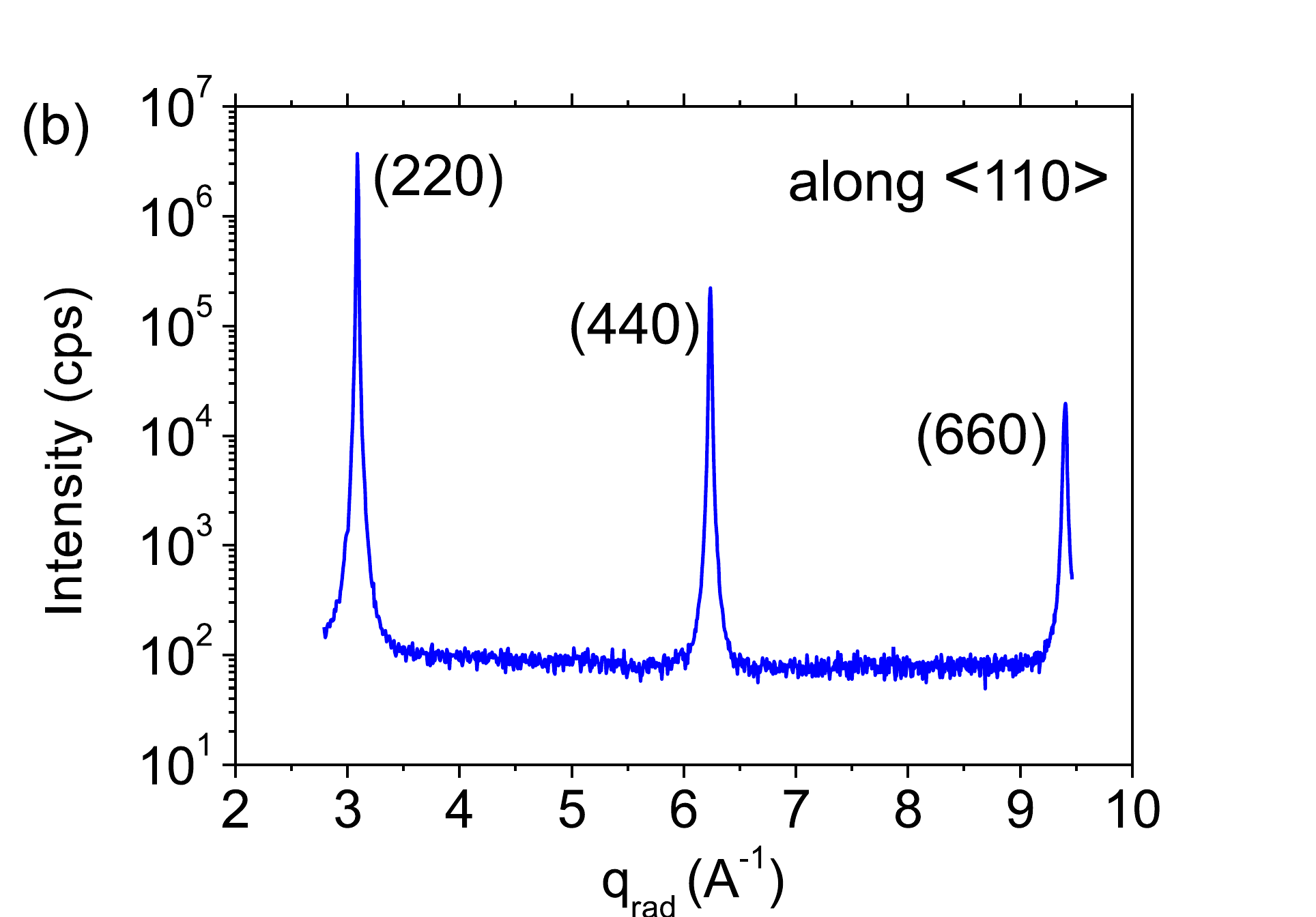}
\includegraphics[width=9.0cm]{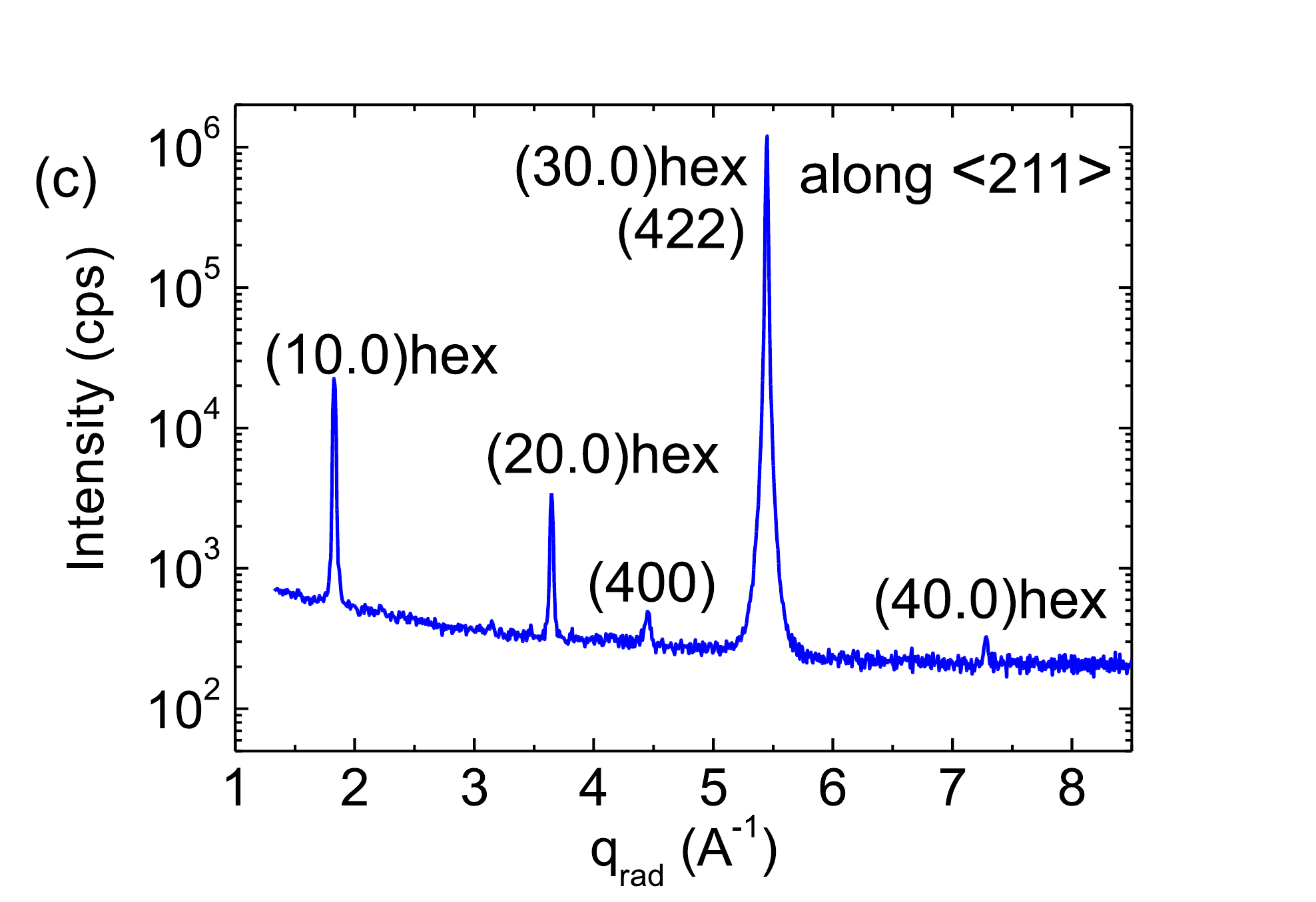}
\caption{(color online) (a) XRD curves of the ($\overline{2}$20) reflection of GaAs/Fe$_{3}$Si core/shell NWs, T$_{S}$~=~200~$^\circ$C. A radial $\omega/2\theta$-scan (thick red line) along the [$\overline{1}$10] direction and an angular $\omega$-scan perpendicular to the [$\overline{1}$10] direction (thin black line), together with their full widths at half maximum (FWHM) are given. Thickness fringes are marked by arrows.
(b) and (c) radial XRD $\omega/2\theta$-scans along $\langle1{1}$0$\rangle$ (b)  and $\langle$211$\rangle$ (c) over a wider range than (a).
}
\label{fig:XRD_curves1}
\end{figure}

% fig.3
\begin{figure*}
\centering
\includegraphics[width=17.5cm]{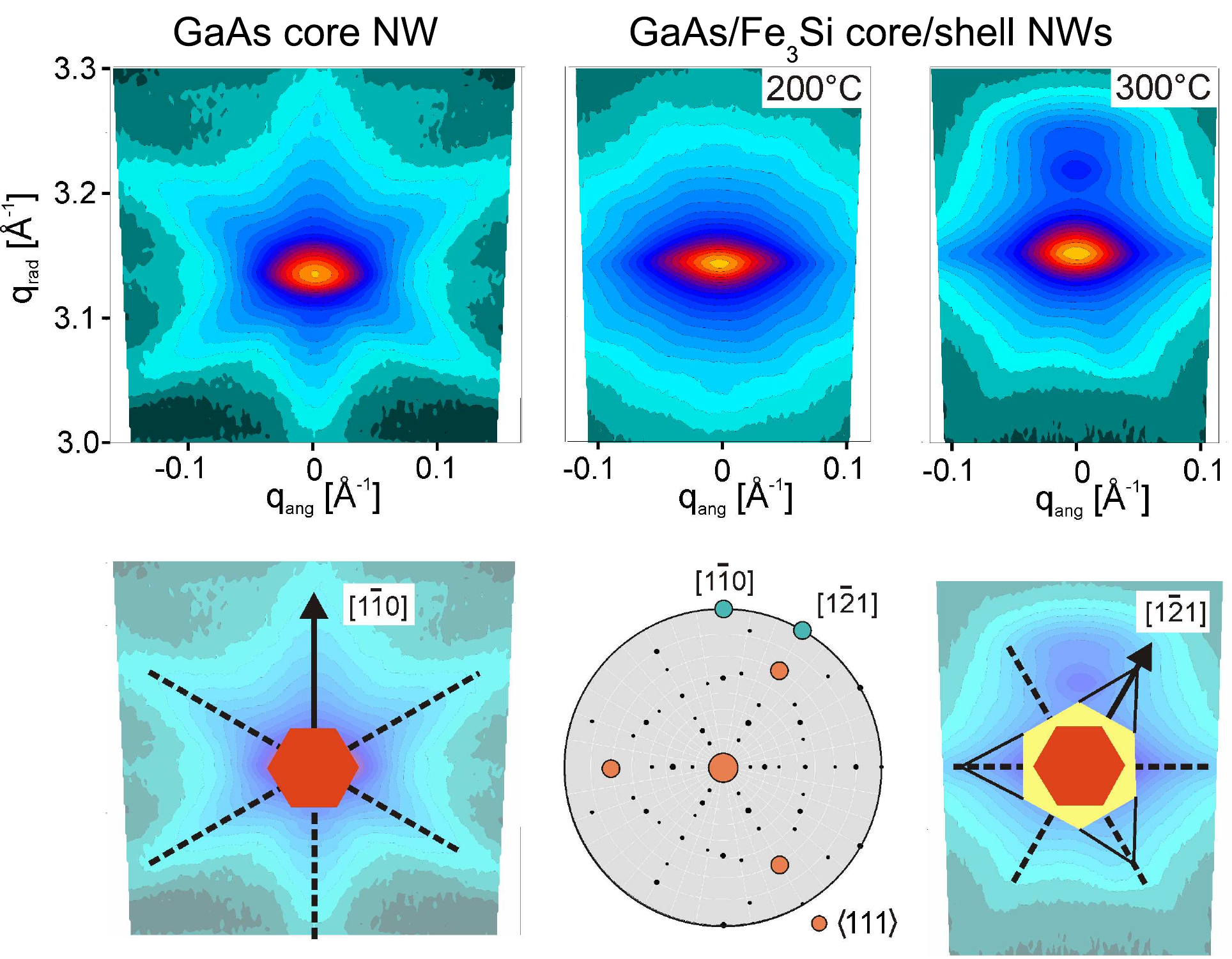}
\caption{(color online) ($\overline{2}$20) in-plane reciprocal space maps of [111]-oriented GaAs NWs (upper left) and GaAs/Fe$_{3}$Si core/shell  NWs (upper center and upper right) grown by molecular beam epitaxy on Si(111). The growth temperatures of the Fe$_{3}$Si shells are given above. The crystallographic directions together with the corresponding sidewalls are sketched below. In the symbolic stereogram (middle) the [1$\overline{1}$0] and [1$\overline{2}$1] directions are marked by blue circles and the different $\{$111$\}$ directions by red circles. The radial direction of the scans is drawn vertically and the angular direction is directed horizontally in the figure.
}
\label{fig:xrd_maps}
\end{figure*}

 % fig. 4
\begin{figure}
%%\centering
\includegraphics[width=9.0cm]{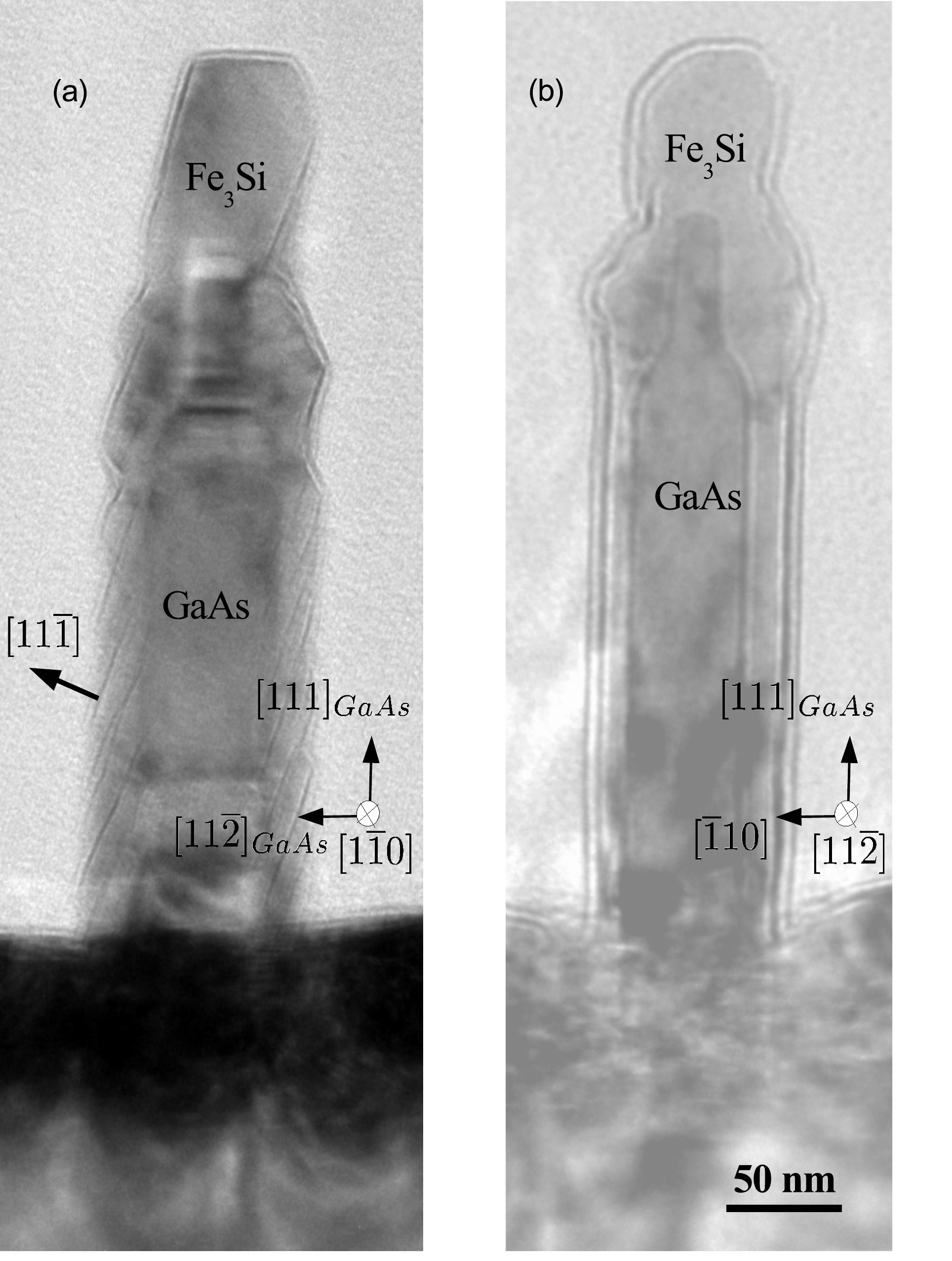}
\caption{Multi-beam bright-field TEM micrographs illustrating the orientations of the Fe$_{3}$Si shell facets with respect to the GaAs cores. The cores and shells are distinguished clearly.  For the [1$\overline{1}$0] direction of the electron beam (a) we see facets corresponding to tilted (11$\overline{1}$) planes as sketched in Fig.~\ref{fig:sketch_facets}.
When the electron beam is running along the [11$\overline{2}$] direction no facets are observed (b). The images are slightly defocused in order to increase the contrast of the surfaces and interfaces.
}
\label{fig:BF}
\end{figure}

An area density of well oriented NWs of $\sim$5$\times$10$^8~$cm$^{-2}$ is found by SEM.  In Fig.~\ref{fig:sketch_facets} SEM top views of a GaAs core NW and a GaAs/Fe$_{3}$Si core/shell NW (T$_S$~=~200~$^\circ$C) are given together with a side--view of a core/shell NW.
 Regular step patterns, facets, are clearly visible on the sidewalls of the NW. At the top of the NW a thinner part is visible. During the last stage of GaAs NW growth no Ga is supplied, and so the remaining Ga in the droplet on top of the NWs is consumed, leading to an elongation of the NW at reduced diameter. In addition Fig.~\ref{fig:sketch_facets} illustrates the orientation of the \{11$\overline{1}$\} shell facets with respect to the cores.
A hexagonal prism symbolizing the GaAs NWs with $\langle$110$\rangle$--oriented sidewalls is drawn  schematically. In addition a tetrahedron depicting the equilibrium shape of Fe$_3$Si crystallites consisting of Fe$_{3}$Si\{11$\overline{1}$\}  planes is shown. The tilted Fe$_{3}$Si\{11$\overline{1}$\} planes form extended shell facets. These facets are intersecting the top Fe$_{3}$Si(111) plane, resulting in triangular features at the top which are visible in the SEM micrograph.

% fig.5
\begin{figure}[!t]
\includegraphics[width=9.0cm]{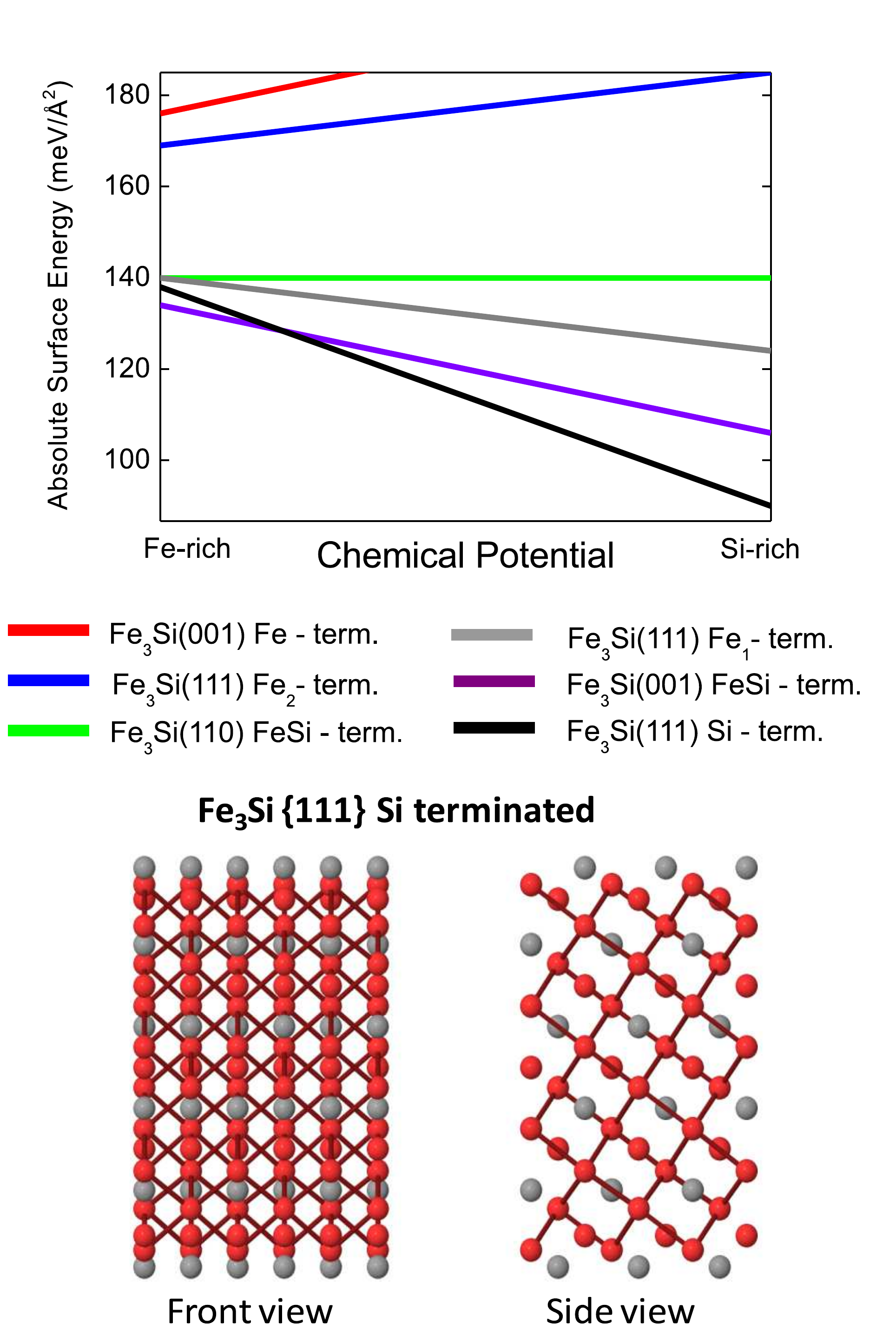}
\caption{(color online) Surface energy of differently terminated (term.) surfaces calculated by DFT in generalized gradient approximation. The structure of the Si--terminated Fe$_{3}$Si\{111\} surface is illustrated below, Fe atoms are symbolized as red balls whereas Si atoms as grey balls.
}
\label{fig:Surface_energy}
\end{figure}

Figure~\ref{fig:XRD_curves1} (a) depicts the XRD curves of the ($\overline{2}$20) reflection of GaAs/Fe$_{3}$Si core/shell NWs, T$_S$~=~200~$^\circ$C.  A radial $\omega/2\theta$-scan (thick line) along the [$\overline{1}$10] direction and an angular $\omega$-scan perpendicular to the [$\overline{1}$10] direction (thin line), together with their full widths at half maximum (FWHM) are given. The radial scan shows thickness fringes (marked by arrows) corresponding to a thickness of 14.3~nm equal to the Fe$_{3}$Si shell thickness (cf.~Fig.~\ref{fig:BF} (b)). The radial scan has a FWHM of 0.08$^\circ$ whereas the angular scan has a larger width of 0.55$^\circ$. The FWHM in angular direction here corresponds to the range of twist of the NWs, as in-plane reflections are used.\cite{Jenichen2011} The range of tilt of GaAs NWs was measured to be near (0.28$\pm$0.1)$^\circ$ using symmetrical out-of-plane measurements.

Figure~\ref{fig:XRD_curves1} (b) and (c) show  XRD curves of GaAs/Fe$_{3}$Si core/shell NWs,  T$_S$~=~200~$^\circ$C. Radial $\omega/2\theta$-scans along (b) the $\langle1{1}$0$\rangle$  and (c) the $\langle$211$\rangle$ directions are given. Scan (b) does not reveal any unexpected maxima due to polycrystalline material whereas scan (c) exhibits additional maxima caused by the wurtzite regions in the GaAs cores. Merely the (400) maximum points to another orientation of some of the crystallites probably caused by additional twins in the cores that are not parallel to the GaAs/Si interface. The detection of the wurtzite regions in the GaAs cores again points to the fact, that our measurements are sensitive for the core/shell NWs only.

Figure~\ref{fig:xrd_maps} demonstrates in-plane reciprocal space maps of the ($\overline{2}$20)-reflection of  GaAs NWs (left) and GaAs/Fe$_{3}$Si core/shell  NWs (center and right) grown by MBE on a Si(111) substrate. The growth temperatures of the Fe$_{3}$Si shells were 200~$^\circ$C (center) and 300~$^\circ$C (right). The crystallographic directions are sketched below. The radial direction ($\omega/2\Theta$) of the scans is vertical and the angular direction ($\omega$)  horizontal. The facets of the  pillar shaped cores with hexagonal cross section are clearly distinguished by streaks in the diffuse scattering pointing along the $\langle$110$\rangle$ directions. Remarkably, in the central map core streaks along the $\langle$110$\rangle$ directions are still visible although they are superimposed by additional streaks along the $\langle$1$\overline{2}$1$\rangle$ directions originating from the facets of the Fe$_{3}$Si shells. In the map shown on the right side only streaks of the shell facets remain. Additional maxima above the main peak indicate chemical reactions between Fe$_{3}$Si and GaAs occurring at T$_S$~=~300~$^\circ$C near the interface similar as those observed for planar films on GaAs.\cite{herfort03,Gusenbauer2011}

Figure~\ref{fig:BF} shows a multi-beam bright-field TEM micrograph, illustrating the orientations of the Fe$_{3}$Si facets of the core/shell NW on Si(111). In Fig.~\ref{fig:BF}(a) the incident electron beam is oriented along [1$\overline{1}$0] in (b) along [11$\overline{2}$]. In (a) we observe facets of the Fe$_{3}$Si shell as tilted black linear contrast steps, in (b) we see smooth interfaces and surfaces. The roughening due to the facetting is not observed along this direction. The visibility of the thickness fringes in Fig.~\ref{fig:XRD_curves1} points to the fact, that (1) we really measure the diffraction signal of the core/shell NWs only, and (2) that the thickness fringes are measured in an ($\omega/2\Theta$)--scan along the [1$\overline{1}$0] direction where the interfaces are smooth.   A growth temperature of 200~$^\circ$C results in this highly perfect Fe$_{3}$Si shell structure.
The (11$\overline{1}$) surface nanofacets are expected to be inclined to the (110) cladding planes of the GaAs cores leading to an increase of the surface area A. There is an non negligible material transport over distances small compared to the NW lengths. On a larger length-scale the Fe$_{3}$Si shell is approximately reproducing the shape of the GaAs core NWs.\cite{jenichen2014} The GaAs cores are usually not free of defects, especially near the Si/GaAs interface and during the late phase of GaAs NW growth planar defects connected to the transition from zincblende to wurtzite segments are found.\cite{Schroth2015} So the straight horizontal contrasts in the GaAs cores in Fig.~\ref{fig:BF}(a) are caused by planar defects leading to alternating zincblende and wurtzite GaAs regions. These planar defects occur only in the vicinity of the GaAs/Si interface and in the region, where the Ga--flux has already been terminated, i.e. near the top of the NWs.

Figure \ref{fig:Surface_energy} shows results of DFT calculations of surface energies of the Fe$_{3}$Si shells for a wide range of chemical potentials and several possible surface terminations. Fe$_{3}$Si surfaces were found to be Si-rich.\cite{Starke2001,Hafner2007} In that case the Si-terminated Fe$_{3}$Si(111) planes are most favorable energetically, even  if geometrical factors, i.e. the inclination angle between GaAs(110) and Fe$_{3}$Si(111) planes of $\psi$~=~35.3~$^\circ$, are taken into account. The (110) surface has an energy of $\gamma_{110}$~=~140~meV$\cdot{\AA}^{-2}$.  Hence the criterion for (111) facet formation is that the (111) surfaces have an energy less than $\gamma_{110}\cdot\cos(\psi)$~=~114.25~meV$\cdot{\AA}^{-2}$. As a result our DFT calculations confirmed that the formation of facets reduces the overall surface energy.

\section{Conclusions}

GaAs core NWs were grown on the oxidized Si(111) surface inside holes of the SiO$_2$ film via the VLS growth mechanism. Then ferromagnetic Fe$_3$Si shells were grown resulting in continuous covering of the cores.
A polycrystal film grew unintentionally between the NWs. We have successfully avoided the XRD signal of the polycrystal film by studying the ensemble of core/shell NWs using x-ray grazing incidence diffraction geometry.
We did not observe additional orientations of the shell with respect to the core, i.e. the shells are pseudomorphic.  Up to a growth temperature of 200~$^\circ$C  additional broadening of the shell diffraction maxima compared to those of the cores is rather limited.
The analysis of the x-ray diffuse scattering revealed the hexagonal cross-section of the GaAs cores. In addition Fe$_{3}$Si nano-facets of the shells cause streaks rotated by an azimuthal angle of 30~$^\circ$. The nano-facets consist e.g. of \{111\} planes tilted around $\langle$011$\rangle$   axes towards the  $\langle$211$\rangle$  direction, i.e. for [111] oriented NWs the ($\overline{1}$11) facets are most pronounced, forming a regular pattern along the GaAs NWs. The \{111\} facets of Fe$_3$Si are formed because under our Si-rich conditions they have the lowest surface energy in agreement with the DFT calculations.  We supported the hypothesis that the nanofacetted  Fe$_3$Si shells found in the present work are a result of VW island growth.\cite{kag09} The role of the surface is more important for NWs than for planar films, the facetted growth of the lattice matched shells is an example for such surface related phenomena. Facetting will also play a role in systems with SK growth mode and strain--driven VW island growth.

\section{Acknowledgement}
The authors thank Claudia Herrmann for her support during the
MBE growth, Doreen Steffen for sample preparation, Astrid Pfeiffer
for help in the laboratory,  Anne-Kathrin Bluhm for the SEM micrographs, Esperanza Luna and Xiang Kong
for valuable support and helpful discussion.
We thank the ESRF in Grenoble for providing beamtime during the experiment HC-1967. We thank Nathalie Boudet and Nils Blanc for their support during the beamtime. The beamtime for some preliminary measurements performed at the PHARAO U-125/2 KMC beamline of the storage ring BESSY II in Berlin is thankfully acknowledged as well. This work was supported in part by the Office of Naval Research through the Naval Research Laboratory's Basic Research Program. Some computations were performed at the DoD Major Shared Resource Center at AFRL.

\section{References}

%%\bibliography{Zitate}

\end{document}